# In situ X-ray diffraction and the evolution of polarization during the growth of ferroelectric superlattices

Benjamin Bein[1], Hsiang-Chun Hsing[1], Sara J. Callori[1,†], John Sinsheimer[1,†], Priya V. Chinta[2], Randall L. Headrick[2] & Matthew Dawber[1]

In epitaxially strained ferroelectric thin films and superlattices, the ferroelectric transition temperature can lie above the growth temperature. Ferroelectric polarization and domains should then evolve during the growth of a sample, and electrostatic boundary conditions may play an important role. In this work, ferroelectric domains, surface termination, average lattice parameter and bilayer thickness are simultaneously monitored using in situ synchrotron X-ray diffraction during the growth of $BaTiO_3/SrTiO_3$ superlattices on $SrTiO_3$ substrates by off-axis radio frequency magnetron sputtering. The technique used allows for scan times substantially faster than the growth of a single layer of material. Effects of electric boundary conditions are investigated by growing the same superlattice alternatively on $SrTiO_3$ substrates and 20 nm $SrRuO_3$ thin films on $SrTiO_3$ substrates. These experiments provide important insights into the formation and evolution of ferroelectric domains when the sample is ferroelectric during the growth process.

[1] Department of Physics and Astronomy, Stony Brook University, Stony Brook, New York 11794-3800, USA. [2] Department of Physics, Cook Physical Science Building, University of Vermont, Burlington, Vermont 05405, USA. † Present addresses: Department of Physics, California State University, San Bernardino, 5500 University Parkway, San Bernardino, California 92407-2393, USA (S.J.C.); Brookhaven National Laboratory, Upton, New York 11973-5000, USA (J.S.). Correspondence and requests for materials should be addressed to M.D. (email: matthew.dawber@stonybrook.edu).

In ferroelectric thin films and multilayers, the polarization is intricately linked to crystal structure, so that strain and electrostatic boundary conditions have considerable impact on the magnitude of the polarization and the arrangement of polarization domains[1–5]. In certain strained ferroelectrics, for example, BaTiO$_3$ (BTO) or PbTiO$_3$ (PTO) grown epitaxially on SrTiO$_3$ (STO), the ferroelectric transition temperature can lie above the growth temperature of the film[6], and thus the electrostatic boundary conditions may also be influential during the growth process. In situ X-ray diffraction during growth of thin films can be a powerful technique for real-time monitoring without interfering with the ongoing process[7–10] and in the case of ferroelectrics can provide an effective probe of ferroelectric polarization through measurement of structural parameters[4–6,11,12].

This study goes beyond thin films and considers the growth of artificially layered ferroelectric superlattices. During the growth of this class of artificial materials, it is desirable to measure the out-of-plane lattice parameters, in-plane lattice parameters, the artificially created superlattice repeat periodicity and the spacing of ferroelectric domains. All of this information can be obtained by performing reciprocal space maps around appropriate Bragg reflections. Here we have developed a scanning technique that makes full use of the available X-ray intensity and area detector technology at X21 at the National Synchrotron Light Source (NSLS) at Brookhaven National Laboratory to provide all of the desired information in substantially less time than it takes to deposit a single unit cell of material. This provides us with an unprecedented ability to monitor the continuous evolution of the polarization and related structural parameters during growth, elucidating the role of electrostatics and strain during the growth of ferroelectric superlattices.

## Results

**Experimental set-up.** Our experiments were performed in an in situ growth chamber at the X21 beamline at NSLS. The chamber is configured as a four-circle diffractometer with control over the $\phi$, $\theta$, $\delta$ and $2\theta$ angles (Fig. 1a,d). Two beryllium windows allow the X-ray beam to enter the chamber, scatter off the sample (which is heated to an appropriate temperature for deposition, in the present work, 650 °C) and exit at the position of the detector, which is a PILATUS-100K area detector. Figure 1a,d shows a schematic of the experimental set-up for scans in the vicinity of the (0 0 1) and (1 0 1) peaks of the substrate. Two shuttered magnetron sources, mounted in an off-axis geometry, enabled the deposition of BTO and STO. Superlattices composed of alternating layers of BTO and STO have been intensively studied in theory and experiment[13–21].

**Growth rate calibration.** To calibrate the growth rates of STO and BTO, the intensity of the signal at the $(0\ 0\ \tfrac{1}{2})$ Bragg position (to minimize bulk Bragg diffraction)[12,22,23] was measured as a function of time. The intensity of this signal provides a measurement of surface roughness. Oscillations in this intensity can be followed in a similar manner to reflection high-energy electron diffraction oscillations[24], though the reflection high-energy electron diffraction technique is not appropriate for the use with magnetron sputtering (due to the magnetic fields present) and does not provide the same amount of structural information as X-ray diffraction can. In the case of layer-by-layer growth, the maxima in the intensity correspond to completed layers, while low-reflectivity signals correspond to incomplete layers. Figure 2 shows this data for BTO, STO and a BTO/STO superlattice. BTO stops growing with well-defined intensity oscillations after two layers, while for STO oscillations can be observed for many layers. In a superlattice with bilayers consisting of two unit cell layers of BTO and six unit cell layers of STO on top of the BTO (2/6 BTO/STO), BTO and STO have intensity oscillations for multiple superlattice bilayers. This allows the growth of a larger total number of BTO layers than in a coherently strained thin film of BTO on STO before the intensity oscillations are no longer observable. Intensity oscillations most likely disappear because of a misfit strain-induced change in growth mode from layer-by-layer growth to island growth. Precise growth rates for the materials are presented in Supplementary Note 1.

**Experimental approach.** Once growth rates had been determined, we investigated the evolution of structural properties related to electrical polarization in ferroelectric superlattices, based on the comparison of the growth of two compositions of BTO/STO superlattices grown with different electrical boundary conditions. The compositions chosen were 2/6 BTO/STO and 1/7 BTO/STO. The reason for choosing these compositions were that the reflectivity measurements showed us that the thin BTO layers within these superlattices could be expected to maintain high quality, while the overall superlattice periodicity of eight unit cells led to ideal positioning of the superlattice peaks on the Pilatus detector. Two samples of each composition were grown, one on a bare STO substrate and the other on top of a ~20-nm SRO film, which provided a conductive boundary condition to the bottom of the sample. The in-plane lattice parameters of the SRO films were constrained to that of the STO substrate.

**Scanning technique.** To be able to rapidly acquire an extensive set of structural parameters during growth, a scanning technique, akin to a time-resolved rotating crystal method[25], was employed. One motor was moved continuously through a given angular range, while the Pilatus detector integrated all the intensity, which reached it during the motion of that motor. This contrasts with a previous rapid root mean squared method developed by Sasaki et al.[26], where a stepped motion of sample and detector were used, and allows us to achieve acquisition times of 15 s as compared with 104 s for their experiment, bring the time required for a map well below the time taken to deposit a single unit cell of material.

For this rapid technique to work, quality of film and substrate needs to be high, in other words they both need to have a sharp rocking curve peak. If they both have a sharp rocking curve peak, the diffraction condition is only met in a very narrow angular range of the scanned motor for each $2\theta$, $\delta$ angle (Fig. 1a,d).

The respective motor movement in each scan corresponds to the movement normally performed during a rocking curve. Therefore, the measured intensity of each pixel is the integrated intensity over the rocking curve at that pixel. This leads to integration over one in-plane momentum transfer direction ($Q_y$), while diffraction information in the two other momentum transfer directions ($Q_x$ and $Q_z$) can be obtained from each camera image.

This technique can be employed for the (0 0 1) reflection by scanning the $\theta$ motor (Fig. 1a), and for the (1 0 1) reflection by scanning the $\phi$ motor (Fig. 1d). During the measurement these are the only motors that are moved. Some further explanation of the method is given in Supplementary Note 2 and Supplementary Fig. 1.

An example of the (0 0 1) data obtained is shown in Fig. 1b. The obtained images can be assembled into continuous movies that allow the observation of the evolution of diffraction features during the growth (Supplementary Movie 1). Our analysis of the data contained in these images is based on two principle lines of points.

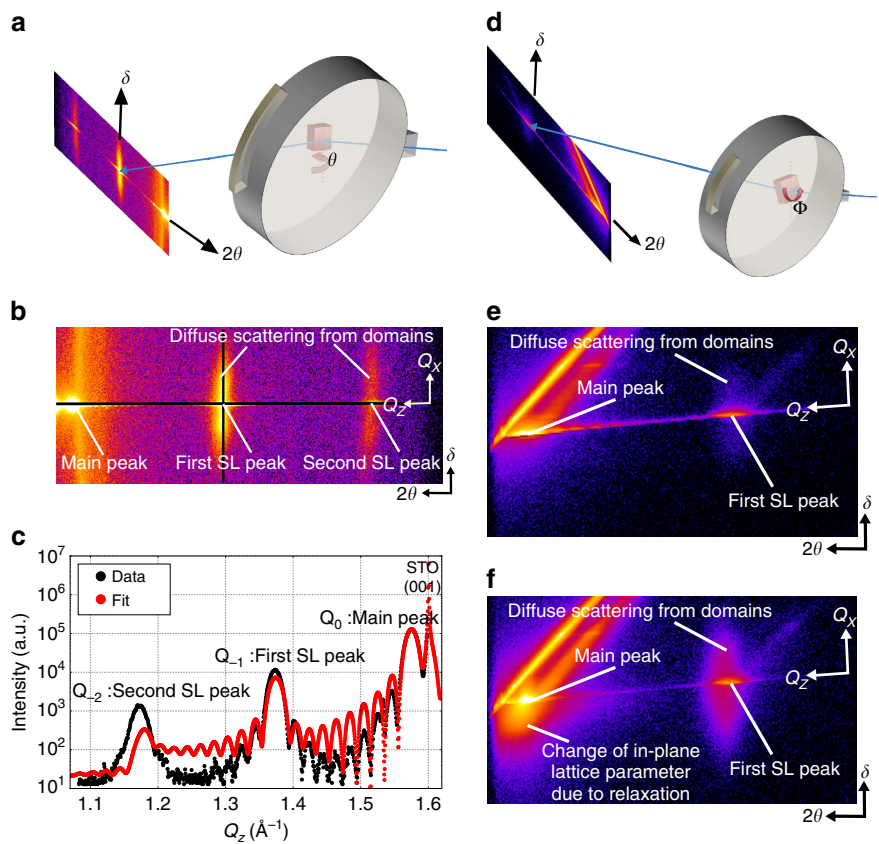

**Figure 1 | Diffractometer angles and example data images.** (**a**) A schematic of the experimental set-up illustrating the angles ($2\theta$, $\theta$ and $\delta$) used for (0 0 l) scans. (**b**) A single exposure of the Pilatus detector during our experiment is shown with the corresponding diffractometer and momentum transfer axis and two black lines of interest. One black line is a horizontal line through the superlattice peaks along the $Q_z$ direction corresponding to a $\theta - 2\theta$ scan. The second black line is a vertical line through the first superlattice peak along the $Q_x$ direction similar to a rocking curve scan through the first superlattice peak. (**c**) Example data (black) along (0 0 l) direction after the growth of 10 bilayers together with a fit (red). (**d**) A schematic of the experimental set-up illustrating the angles ($2\theta$, $\Phi$ and $\delta$) used for (1 0 l) scans. (**e**) Image of a single scan around (1 0 l) after 10 bilayers is shown with the corresponding diffractometer and momentum transfer axis. (**f**) Image of a single scan around (1 0 l) after 30 bilayers with the corresponding diffractometer and momentum transfer axis. In the Pilatus detector images shown, intensity is represented on a logarithmic flame colour scale (from low to high intensity, black, violet, red, orange, yellow and white).

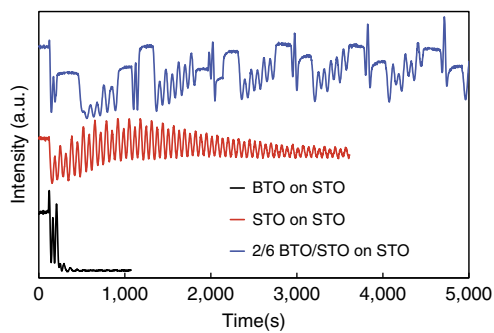

**Figure 2 | Growth rate calibration using the anti-Bragg peak.** Intensity (shifted so that several curves can be seen on the same graph) of the (0 0 $\frac{1}{2}$) anti-Bragg peak for BTO on STO (black), STO grown on STO (red) and BTO/STO 2/6 superlattice grown on STO (blue). It can be seen that the BTO stops growing in a smooth way after two layers. The STO on the other hand continues to grow smoothly in a layer-by-layer mode for many layers. The BTO/STO superlattice continues to show intensity oscillations for each BTO double layer. STO layers in the superlattice help maintain the BTO growth in a smooth layer-by-layer mode.

The first line is along the $Q_z$ direction, which is a horizontal line in Fig. 1b. $Q_z$ is the momentum transfer in the $z$ or out-of-plane direction. A plot along this line together with a fit to the data and basic superlattice attributes are shown in Fig. 1c.

The second line of interest in Fig. 1b is a vertical line through the first superlattice diffraction peak. It is along the $Q_x$ direction and gives information about the polarization domain periodicity in the superlattice. $Q_x$ is the momentum transfer in the $x$ direction, which is one of the in-plane directions.

**Data analysis.** Much can be learned about the superlattice from the data on the horizontal line along (0 0 l) ($Q_z$ direction). For instance, the position of the main, first and second superlattice peak can be tracked during the growth. The average out-of-plane lattice parameter, $\bar{c}$, can be approximated directly from the measured position of the main superlattice peak ($\bar{c} = \frac{2\pi}{Q_0}$, $Q_0$ is the momentum transfer at the main superlattice peak)[27] and the result is shown in Fig. 3a.

This approach is reliable after the superlattice has a total thickness of 30 nm, or in this case, 10 bilayers. Until that thickness is reached the superlattice peak intensity is weak compared with the tail of the substrate peak, which affects the apparent value of the lattice parameter of the film. On samples with a SRO bottom, electrode interference with the diffraction from the thin-film electrode makes it even harder to determine $\bar{c}$ accurately by direct measurement of the film peak intensity. As we discuss later, fitting the data using kinematic X-ray theory can overcome these problems.

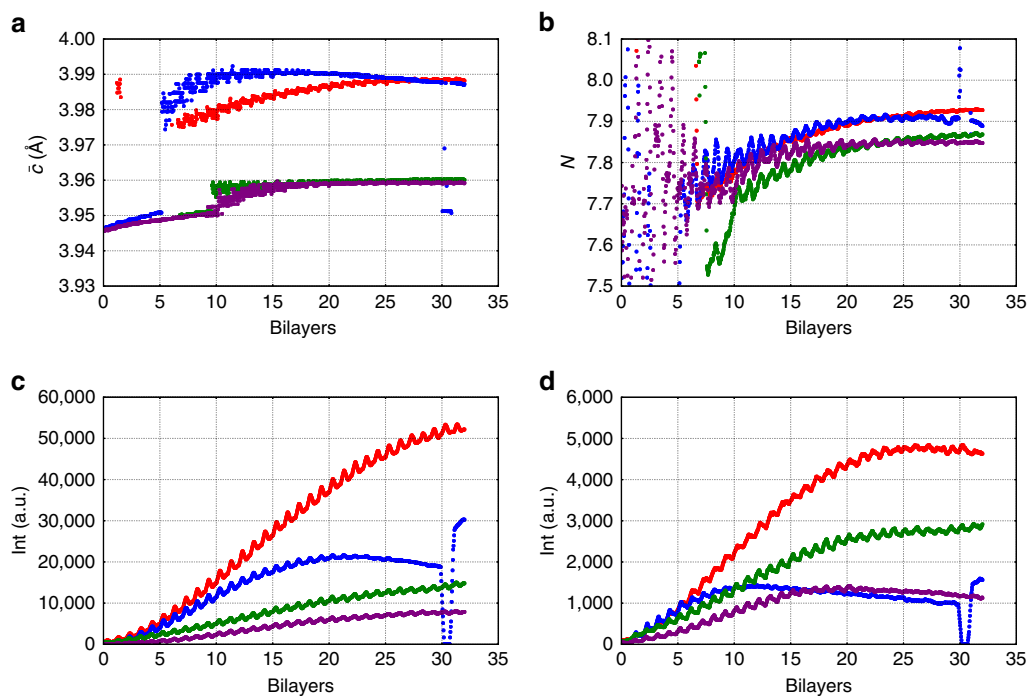

**Figure 3 | Evolution of superlattice parameters and peak intensities.** The figure shows (**a**) average out-of-plane lattice parameter $\bar{c}$, (**b**) Number of layers per bilayer $N$, (**c**) first and (**d**) second superlattice peak intensity plotted versus the number of bilayers. In each panel, the 2/6 BTO/STO on STO is plotted in blue, 2/6 BTO/STO on SRO in red, 1/7 BTO/STO on STO in violet and 1/7 BTO/STO on SRO in green. From (**b**) the plot of the number of unit cell layers per bilayer, $N$, it can be seen that all the superlattices are grown close to the desired eight unit cell layers per bilayer. The earlier reduction of the superlattice quality of the samples grown on plain STO (blue and violet) can be seen in the decay of the first and second superlattice peak intensities (plots **c** and **d**), as well as in the reduction of $\bar{c}$ with increasing thickness in **a**. A larger $\bar{c}$ for the 2/6 BTO/STO (blue and red) than the 1/7 BTO/STO superlattice (violet and green) is also seen in **a** arising from the larger-volume fraction of BTO.

The number of unit cell layers per bilayer $N$ can be calculated from the bilayer thickness $\Lambda$, and the average out-of-plane lattice parameter $\bar{c}$, as $N = \frac{\Lambda}{\bar{c}}$. $\Lambda$ and $\bar{c}$ can be expressed in terms of the superlattice peak positions shown in Fig. 1c, $Q_0$ is the momentum transfer of the main peak and $Q_{-1}$ is the momentum transfer of the first negative superlattice peak. So $\bar{c} = \frac{2\pi}{Q_0}$, $\Lambda = \frac{2\pi}{Q_0 - Q_{-1}}$ and $N = \frac{\Lambda}{\bar{c}} = \frac{Q_0}{Q_0 - Q_{-1}}$.

Figure 3b shows the evolution of $N$ during growth and reveals that the grown superlattice has ~7.9 unit cells per bilayer, which is close to the desired eight unit cell layers per bilayer.

By tracking the first and second superlattice peak intensity, the quality of the superlattice can be monitored. A high-quality superlattice will have an increasing peak intensity with increasing thickness (it is expected that the intensity should scale with the square of the film thickness in an ideal case), while the intensity will be lower than the expected value if the superlattice loses its high quality. Figure 3c,d shows plots of the first and second superlattice peak intensities against the superlattice thickness.

The data show that samples grown on SRO bottom electrodes maintain excellent quality during the whole growth up to 100 nm, while samples grown on plain STO substrates start losing superlattice peak intensity after 30 nm. The most likely cause of this change in intensity is in-plane relaxation of the superlattice. This can be demonstrated by example images of scans around the (1 0 1) peak. These peaks contain not only information about the out-of-plane lattice parameter but also about the in-plane lattice parameters. The scan after 30 bilayers (Fig. 1f) shows a feature under the main superlattice peak. This feature is an indication of an in-plane lattice parameter change and it is not present for thinner superlattices (see Fig. 1e, same superlattice with 10 bilayers grown).

### Discussion

To extract reliable values of structural parameters from the experimental data, appropriate fitting approaches (described in Methods) had to be applied. This is particularly important for very thin films or superlattices with a small number of bilayers, where the intensity of scattering from the substrate is greater than the intensity of the superlattice peaks, so that clearly defined peaks are not observed. However, by fitting model parameters so that the calculated diffraction patterns match the measurements, one can extract the desired information even for very thin films, well before clear peaks become visible.

It became apparent from the fitting that the termination of our superlattices changes during the growth process. By comparing different simulated terminations with the experimental data (Supplementary Note 3; Supplementary Fig. 2), we found that after one bilayer was deposited, the fits to the diffraction data indicated the STO layers had a SrO termination and the BTO layers had BaO termination. This is interesting considering that the initial termination of the substrates was known to be TiO$_2$. First principles calculations do indicate that the BaO termination is energetically more favourable than TiO$_2$ in ultra-thin BTO thin films at this strain condition[28], which may be the driving force for this change of termination in the superlattice.

Known elastic constants for BTO[29] and STO[30] were used to calculate the expected value for $\bar{c}$ in the absence of polarization (black curve) and compared with our fitted values in Fig. 4a,b,d,e. It is obvious that the measured values are higher than a paraelectric sample would display. This is explained by the existence of ferroelectric polarization, which is coupled with $\bar{c}$ (refs 3,27).

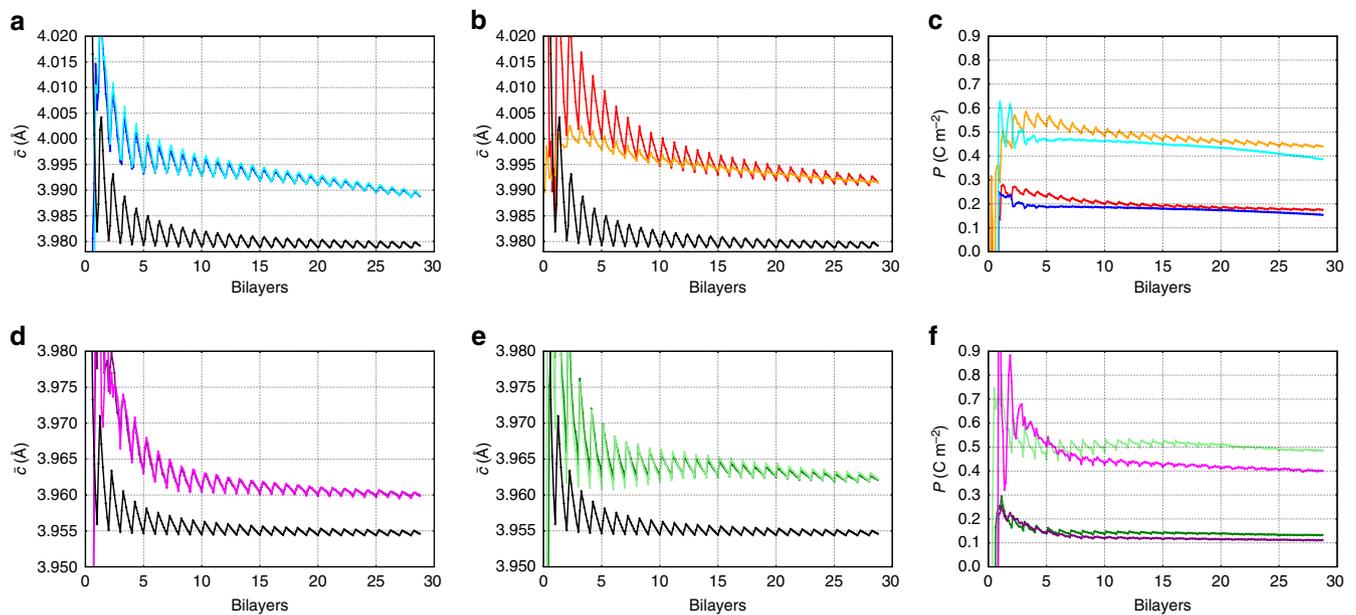

**Figure 4 | Evolution of fitted out-of-plane parameter and polarization.** Fitted $\bar{c}$ values for the (**a**) 2/6 BTO/STO sample grown on STO (blue: homogeneous polarized model, turquoise: only BTO-polarized model), (**b**) 2/6 BTO/STO sample grown on SRO (red: homogeneous polarized model, orange: only BTO-polarized model), (**d**) 1/7 BTO/STO sample grown on STO (violet: homogeneous polarized model, pink: only BTO-polarized model), (**e**) 1/7 BTO/STO sample grown on SRO (green: homogeneous polarized model, lime: only BTO-polarized model) together with expectations from elastic theory (black curves) for non-polar samples. Like the raw data, the fits show that samples grown on SRO have a larger value for $\bar{c}$, which is explained by a larger polarization within the elastic theory. The corresponding polarization for the two limiting models (the homogenous polarization scenario in which BTO and STO carry the same polarization and the scenario only BTO is polarized) are shown in (**c**) for the 2/6 BTO/STO superlattices (blue: grown on STO, homogeneous polarized model, turquoise: grown on STO, only BTO-polarized model, red: grown on SRO, homogeneous polarized model, orange: grown on SRO, only BTO-polarized model) and (**f**) for the for the 1/7 BTO/STO superlattices (violet: grown on STO, homogeneous polarized model, pink: grown on STO, only BTO-polarized model, green: grown on SRO, homogeneous polarized model, light green: grown on SRO, only BTO-polarized model). The difference of the 2/6 BTO/STO superlattices from the elastic theory value is more than twice as big as that of the 1/7 BTO/STO samples, implying that the out-of-plane polarization of the 2/6 BTO/STO samples is larger than in the 1/7 BTO/STO samples.

Furthermore, $\bar{c}$ is larger for the samples grown on a SRO bottom electrode, suggesting that a superlattice grown on a SRO bottom electrode is more polar. After the samples had been cooled and removed from the growth chamber piezo force microscopy (shown in Supplementary Note 4 and Supplementary Fig. 3) was successfully performed on the samples to confirm that they are indeed ferroelectric. On some of the samples, the lattice parameter was monitored during cooling and no evidence of a ferroelectric phase transition between room temperature and 650 °C was detected (Supplementary Fig. 4).

Some care needs to be taken with the use of elastic constants and lattice parameters for stoichiometric materials in this kind of analysis. For example, when Zubko et al.[31] investigated the lattice parameters of PTO/STO superlattices at room temperature, they used a STO lattice parameter of 3.92 Å, rather than the typical value of 3.905 Å. This increased lattice parameter was justified based on measurements of a STO thin film and is associated with imperfect stoichiometry of the STO film[32]. However, in the case of our experiments, when a $\theta - 2\theta$ scan around the (0 0 1) substrate peak for the STO film grown in Fig. 2b was performed immediately after the growth had finished, it was nearly indistinguishable from that performed on the substrate before deposition. Our growth parameters are somewhat different from those used by Zubko et al.[31], who used a pressure of 0.18 torr and a temperature of 520 °C, compared with the pressure of 0.025 torr and 650 °C that we used.

Two limiting cases are considered for the possible distribution of polarization between the layers in the superlattice. The first case is extremely strong coupling between layers, leading to a homogeneous polarization model. In this model, polarization can be considered homogeneous throughout the superlattice, with both materials (BTO and STO) having the same polarization value $P^{17,33}$. This results in a polar dependence for both the out-of-plane lattice parameter of BTO $c_{BTO}(P)$ and the out-of-plane lattice parameter of STO $c_{STO}(P)$, according to conventional strain polarization coupling relationships. In the other extreme, in which layers are essentially decoupled, the ferroelectric material, BTO, is polarized and the dielectric material, STO, has no polarization, that is, an only BTO-polarized model. This time only $c_{BTO}(P)$ depends on the polarization and $c_{STO}$ is independent of the polarization. The two models are illustrated in Supplementary Fig. 5 and further explained in Supplementary Note 5. Most likely, the actual distribution of polarization will be somewhere between these two extremes[34,35]. For both of these models, the polarization value was used as a fitting parameter to produce the lattice parameters that go into the X-ray diffraction equations, and varied to obtain the best fit to the experimental diffraction pattern. The evolving diffraction patterns and fits for all four samples can be seen in Supplementary Movies 2–5. Following the fitting, $\bar{c}$ is extracted from the lattice parameters that produced the best fit. The results are shown in Fig. 4a,b,d,e. Our measured values can be compared with simple calculations we performed based on the thermodynamic potential of Li et al.[36], which was found to perform well for BTO grown on $DyScO_3$ and $GdScO_3$ (ref. 29). The cubic lattice parameter of bulk BTO single crystal at 650 °C is 4.025 Å (measured by Choi et al.[29]), while that of STO (measured directly in this experiment) is 3.93 Å. The mismatch strain at the growth temperature for BTO on STO is therefore −2.3%. The thermodynamic potential of Li et al. predicts that BTO at this strain and temperature should be ferroelectric, with a polarization of $0.24\,C\,m^{-2}$. This value is much closer to that of the homogeneous polarization model ($P \sim 0.2\,C\,m^{-2}$). For the only BTO-polarized model, the values of polarization ($P \sim 0.45\,C\,m^{-2}$) are implausibly high. The development of polarization during the growth for both models is shown in Fig. 4c,f.

The initial downward trend in $\bar{c}$ comes from the fact that the first layer deposited is BTO and so the lattice parameter oscillates upwards when the composition of the superlattice is BTO rich, with the $\bar{c}$ value for the superlattice composition acting as a lower bound. As more material is added the degree to which the composition is BTO rich decreases and the mean value of $\bar{c}$ is reduced. The observed oscillation in the out-of-plane lattice parameter has two sources, the main one is that BTO has a larger unit cell than STO, but there are also changes due to the evolution of polarization as the film grows. As each BTO layer is deposited, the polarization increases a little and then its decreases again as the next STO layer is deposited. For samples grown on SRO, this sawtooth-like oscillation of the polarization is more pronounced.

Another indication of ferroelectric polarization is diffuse scattering along the vertical line ($Q_x$ direction) through the first superlattice peak in Fig. 1b. This is a sign of periodic in-plane features, which are typically associated with a polarization stripe domain structure in ferroelectric materials. We note that because of the integration in $Q_y$ that occurs in our method, the peaks in the diffuse scattering associated with domains are somewhat smeared out to how they would appear in a conventional X-ray diffraction scan. Nevertheless, the positions of the peaks can be used to track evolution in the domain size as the film grows quite accurately. Figure 5a shows the diffuse scattering together with the fit to the data. Evolution of the diffuse scattering and fits for each of the four samples can be seen in Supplementary Movies 6–9.

To analyse the diffuse scattering, we assumed a Gaussian distribution of the stripe domain size and a Lorentzian intensity distribution for the superlattice peak. The position of the fitted Gaussian is the average domain size of the superlattice and is plotted against the total superlattice thickness in Fig. 5c.

To highlight the shape of the diffuse scattering and its change with increasing superlattice thickness, the data were rescaled in Fig. 5b. The rescaling was done by dividing the data by the peak diffuse scattering intensity from the fit. The early stage (after 7 bilayers) and the late stage (after 25 bilayers) of the 2/6 BTO/STO superlattice grown on STO are compared in Fig. 5b. This comparison reveals that domain size remains fairly constant despite the thickness of the film changing substantially, which we saw consistently for all of the superlattices (Fig. 5c).

The $\bar{c}$ measurements suggest that polarization is rather homogeneous with similar values of polarization in the BTO and STO layers. In this scenario, it might be expected that the domain size evolves during growth according to the Kittel Law[37], which predicts that the domain wall period should scale as $t^{\frac{1}{2}}$, as the thickness of the film, $t$, increases. Such behaviour is not observed, but it is noticeable that the domains are substantially bigger for the 2/6 BTO/STO sample grown on SRO, which aligns with the expectation that the screening provided by a metallic bottom electrode should reduce the need for domain formation.

In experiments on PTO thin films where scaling of domain size with thickness was observed[4], the domains form once a film of a given thickness has been grown in the paraelectric phase and subsequently cooled. By contrast, in our experiments, domains form at some point in the growth process and the film thickness continues to increase after this has occurred. Our results suggest that domain size becomes locked very early in the growth of the superlattice, and so while electrostatic boundary conditions influence domains in the very early stages of growth they do not change very much after that. The domains are larger than those expected for a single BTO layer (which should be smaller than 5 nm in size[38]), but do not evolve to the size one would expect for the full superlattice thickness.

A locked domain period does not imply that the polarization state of the superlattice is fixed[39–42], and the oscillations we observe in the polarization required to fit the structural data would suggest that the polarization is actually in a continuous state of evolution as the film grows.

An increase in domain size requires domain walls to annihilate so that larger domains can be created and although the system would benefit energetically from an increase in domain size, there needs to be sufficient thermal motion of domain walls for annihilation events to take place, and this does not appear to occur here. Paruch et al.[43,44] observed that thermal roughening of domain walls in PZT thin films is on the order of 1–2 nm in the temperature region used for growth in these experiments, well below the measured domain sizes here ($\sim 20$–$30$ nm) and 90° domain walls in bulk BTO also seem to be stable against temperature induced annihilation until very close to the ferroelectric–paraelectric phase transition[45–47]. Our finding that domain size in ferroelectric superlattices can be influenced by electrostatic boundary conditions during growth and locked in early in the deposition process can provide a route for engineered domain sizes precisely targeted towards particular applications[48].

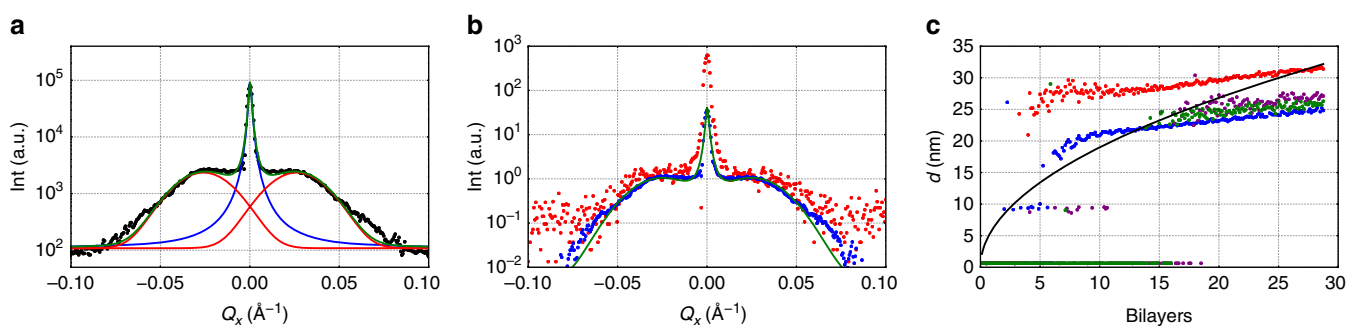

**Figure 5 | Ferroelectric domain fitting and evolution.** (**a**) A plot through the diffuse scattering region along the $Q_x$ line together with a fit (green) to the data (black). To fit the diffuse scattering, we assumed a Gaussian distribution of domain sizes, these are illustrated in red. For the main peak, we used a Lorentzian distribution, which is shown in blue. (**b**) Data from early in the growth (after 7 bilayers in red) and later (after 25 bilayers in blue) were rescaled by dividing these with the peak diffuse scattering intensity of the fit and are shown together with a rescaled fit (green). (**c**) The average domain size plotted against the sample thickness (2/6 BTO/STO on STO in blue, 2/6 BTO/STO on SRO in red, 1/7 BTO/STO on STO in green and 1/7 BTO/STO on SRO in violet). One can see that the square root relationship (black) between domain size and film thickness typically expected for ferroelectric systems does not describe our data well.

To summarize our work, we have developed a scanning method that allows X-ray diffraction reciprocal space maps to be acquired very rapidly during the growth of thin films. Using this technique, we have looked at the important case of a highly strained ferroelectric superlattice that acquires a ferroelectric polarization during the growth process. We have shown that in this scenario electrical boundary conditions have significant impact on the polarization domain structure of the material, and that this domain structure becomes locked in very early in the growth, in contrast to other ferroelectric thin films where the domain structure emerges on cooling from the elevated temperatures used for deposition and depends strongly on the thickness of the sample.

## Methods

**Details about the experimental set-up.** The experiments presented here were performed at the NSLS beam line X21 using an X-ray energy of 10 keV. The energy was selected by a Si(111) monochromator and the beam is focused at the experiment by a bendable cylindrical mirror. The spot size on the sample was $\sim 0.5 \times 0.5$ mm. The experiments were performed in an in situ growth chamber with temperature, pressure and atmospheric control. The chamber is configured as a four-circle diffractometer with control over the $\phi$, $\theta$, $\delta$ and $2\theta$ angles (Fig. 1a,d). Two beryllium windows allow the X-ray beam to enter the chamber, scatter off the sample and exit at the position of the detector, a PILATUS-100K area detector. All films and superlattices grown in this experiment were grown by off-axis magnetron sputtering and under the same growth conditions. The pressure was kept at 0.025 torr with an oxygen/argon ratio of 7:16. The growth temperature was 650 °C. The radio frequency power applied to the magnetron sources was 30 W.

**Substrate and substrate treatment before deposition.** The materials that were deposited in the chamber in these experiments were BTO and STO. In all experiments, high-quality STO substrates from CrysTec GmbH with miscut angles $\sim 0.15°$ and single unit cell steps of $\sim 0.4$ nm were used. The substrates had been etched in buffered hydrofluoric acid and annealed by the vendor to ensure a $TiO_2$ surface termination. Some experiments were performed with deposition directly on to the STO substrates, which provides an insulating boundary condition. When this was done the substrates were annealed for 20 min at 750 °C in a 0.05 mtorr oxygen atmosphere directly before deposition to improve the quality of the surface. Other experiments were performed with deposition onto SRO bottom electrodes to provide a conductive boundary condition. The SRO films used were 20-nm thick and grown before the in situ experiments in an off-axis magnetron sputtering chamber at Stony Brook University. The SRO films were also atomically flat with single unit cell steps of 0.4 nm, and it was verified by X-ray diffraction before deposition that these films were epitaxially constrained to the STO substrate and had the same in-plane lattice parameter as the substrates.

**Fitting method.** The starting point of the fitting is the diffraction equation of kinematic X-ray diffraction, which can be found in textbooks[25,49].

In this paper the (0 0 l) crystal truncation rod is investigated with quantitative fits so that for the fits only $Q_z$ is non 0. In this case, the diffraction amplitude $A$ from a thin film of $N$ unit cells can be approximated by:

$$A \propto \sum_{n=0}^{N} F(Q_z) \cdot \exp(-i \cdot Q_z \cdot c \cdot n - \epsilon \cdot n) = F(Q_z) \frac{1 - \exp(-i \cdot Q_z \cdot c \cdot N - \epsilon \cdot N)}{1 - \exp(-i \cdot Q_z \cdot c - \epsilon)}, \quad (1)$$

where $F(Q_z)$ is the structure factor of the film material, $\epsilon$ is the absorption coefficient, $c$ is the out-of-plane lattice parameter and $N$ is the number of unit cell layers. To calculate the diffraction pattern of the superlattice, one needs to add up the diffraction amplitude from all films and multiply it by the phase shift to the surface of the sample. For the whole superlattice this leads to:

$$A_{total} = A_{sub} \cdot \exp(-i \cdot Q_z \cdot (t_{el} + t_{SL})) + A_{el} \cdot \exp(-i \cdot Q_z \cdot t_{SL}) + A_{SL}$$
$$A_{SL} = \sum_{l=0}^{N_{SL}} (A_{STO} + A_{BTO} \cdot \exp(-i \cdot Q_z \cdot t_{STO})) \cdot \exp(-i \cdot Q_z \cdot (t_{STO} + t_{BTO}) \cdot l), \quad (2)$$

where $t_{el}$, $t_{SL}$ as the total thickness of the electrode, superlattice, respectively. $N_{SL}$ is the number of bilayers in the superlattice. $t_{STO}$ and $t_{BTO}$ are the thickness of the STO and BTO layer/film in one bilayer respectively. $A_{total}$, $A_{sub}$, $A_{El}$, $A_{STO}$ and $A_{BTO}$ are the total, substrate, electrode, STO layer/film within each bilayer and BTO layer/film within each bilayer diffraction amplitude, respectively, and can be calculated using equation (1). $A_{SL}$ is the superlattice diffraction amplitude. A standard gaussian surface roughness term of $\exp(-\sigma^2 \cdot (Q_z - Q_{001})^2)$ was used with $\sigma$ as the root mean squared roughness of the surface and $Q_{001} = \frac{2\pi}{c}$ as the momentum transfer of the (001) peak of the material. This surface roughness was applied to each part of the sample ($A_{sub}$, $A_{el}$, $A_{STO}$ and $A_{BTO}$), where $Q_{001}$ was chosen according to the material ($c_{sub}$ out-of-plane lattice parameter of STO, $c_{el}$ out-of-plane lattice parameter of SRO, $c_{STO}$ out-of-plane lattice parameter of STO and $c_{BTO}$ out-of-plane lattice parameter of BTO). Non-integer film thicknesses are achieved by averaging over integer films.

### Acknowledgements
This work was supported by NSF DMR1055413. Development of the in situ growth facility used in this work was supported by NSF DMR0959486. P.V.C. and R.L.H. were supported by US Department of Energy, Office of Science, Office of Basic Energy Sciences, under Contract No. DE-FG02-07ER46380. A. DeMasi, M.H. Yusuf, H. Yusuf, J.G. Ulbrandt, S. LaMarra and C. Nelson are thanked for their assistance with the experimental work. Use of the NSLS, Brookhaven National Laboratory, was supported by the US Department of Energy, Office of Science, Office of Basic Energy Sciences, under Contract No. DE-AC02-98CH10886. M.D. acknowledges support from the Donostia International Physics Center and CIC Nanogune, which enabled helpful discussions with P. Aguado-Puente and E. Artacho.

### Author contributions
All authors participated in the experimental work at NSLS. B.B. was chiefly responsible for the subsequent processing and analysis of the data. The experimental apparatus was designed, assembled and integrated in to the X21 beamline by J.S., R.L.H. and M.D. M.D. devised the experimental methods and directed the project. The manuscript was prepared by B.B. and M.D. with contributions from the other authors.

### Additional information
**Supplementary Information** accompanies this paper at http://www.nature.com/naturecommunications

**Competing financial interests:** The authors declare no competing financial interests.

**Reprints and permission** information is available online at http://npg.nature.com/reprintsandpermissions/

**How to cite this article:** Bein, B. et al. In situ X-ray diffraction and the evolution of polarization during the growth of ferroelectric superlattices. Nat. Commun. 6:10136 doi: 10.1038/ncomms10136 (2015).